# Hydrodynamic fluctuations of liquids with internal rotation inside a pore

T. Yu. Tchesskaya[a], V. Lisy[b] and A. V. Zatovsky[c]

[a]Information Technology Department, Odessa State Ecological University, Lvovskaya 15, 65016 Odessa, Ukraine[*]

[b]Biophysics Department, P.J. Safarik University, Jesenna 5, 041 54 Kosice, Slovakia[†]

[c]Theoretical Physics Department, I.I. Mechnikov Odessa National University, Dvoryanskaya 2, 65026 Odessa, Ukraine[‡]

The correlation theory of the thermal hydrodynamic fluctuations of liquids with internal spin within a spherical cavity is developed. For the hydrodynamic fields of such a liquid, linearized equations with random thermal sources are used. For the time dependent amplitudes of the expansion of the fields in the basis functions of the vector Helmholtz equations a set of Langevin equations is obtained that allowed us, with the help of the fluctuation-dissipation theorem, to find the spectral densities of the correlation functions of the amplitudes. The spectra of the fluctuations of hydrodynamic fields are found in the form of a superposition of spectral amplitudes of the expansion with weights expressed as bilinear combinations of the proper coordinate functions and their derivatives. The correlation functions of the field of the translational velocity and internal spin of the liquid in one spatial point are analyzed. They have a sense of local coefficients of the translational and rotational diffusion coefficients of the Lagrange particle. We represent them as a series in spherical Bessel functions and roots of transcendental equations following from the boundary conditions on the sphere. The results demonstrate that the size of the pore essentially affects the character of the hydrodynamic fluctuations of the liquid.

The equilibrium and dynamical properties of liquids in small restricted volumes significantly differ from those in large volumes [1-4]. The experiments show that the shear viscosity of some liquids as a phenomenological parameter loses its physical meaning of a material constant and near a solid surface becomes an effective quantity that depends on the distance from the wall [5]. In the classical Navier-Stokes hydrodynamics there is no difference between the internal friction of the liquid shells with one another and the external friction on the solid boundary. The boundary must however cause an anisotropy of the properties of the liquid being in contact with it. The ordering of the liquid can be connected either with the orientation of the molecules at the boundary or with the change of their translational and rotational mobil-ity. The account for such effects is possible in the following way: we shall assume that the liquid, in addition to the density of the kinetic moment connected with the translational motion of the molecules contains an intrinsic moment of momentum – the spin moment $\vec{s} = \vec{m}/I$ that is due to the rotation of the molecules ($I$ is the volume density of the inertia moment of the molecules). For the first time the necessity of the account for the rotation of the molecules in liquids has been shown by Ya.I. Frenkel [6]. Later [7,8], the hydrodynamics of liquids with additional degrees of freedom has been built. The derivation of the extended set of equations of such a liquid that in addition to the shear and bulk viscosity contains the rotational viscosity, has been described in detail in the work [8]. Much of the experimental data given in Refs. [1,5] are well explained on the basis of the system of hydrodynamic equations taking into account the local rotational velocity and appropriate boundary conditions in which a new kinetic coefficient is present: the boundary viscosity.

[*] email: tatyana@odessa.zzn.com
[†] email: lisy@kosice.upjs.sk
[‡] email: avz@dtp.odessa.ua

In the present contribution the thermal hydrodynamic fluctuations are studied in the case when for the liquid besides the velocity field of translational motion the field of the local rotational velocity is taken into account, and the liquid itself is placed in a rigid spherical cavity.

The linearized equations of motion for a compressible fluid with internal rotations are

$$\rho_0 \frac{\partial \vec{v}}{\partial t} = -c^2 \nabla \delta\rho + (\eta + \frac{\eta_r}{2})\Delta\vec{v} + (\xi + \frac{\eta}{3} - \frac{\eta_r}{2})\nabla \mathrm{div}\,\vec{v} + \eta_r \mathrm{rot}\,\vec{s} + \vec{f}_v, \qquad (1)$$

$$\rho_0 \frac{\partial \vec{s}}{\partial t} = -\mu\,\mathrm{rot}\,\mathrm{rot}\,\vec{s} + \eta_r\left(\frac{1}{2}\mathrm{rot}\,\vec{v} - \vec{s}\right) + \vec{f}_s, \qquad \vec{s} = \frac{\vec{m}}{I},$$

where $\vec{v}$ and $\vec{s}$ are the translational and rotational velocity, respectively, $\rho_0$ is the equilibrium mass density of the liquid, $c$ is the velocity of sound, $\vec{f}_{v,s}$ are external fluctuational forces, and $\mu$ is the diffusion coefficient of the internal momentum. The coefficients $\eta_r$, $\eta$, and $\xi$ denote the rotational, shear, and bulk viscosity, respectively.

The closed system of hydrodynamic equations includes also the continuity equation and the assumption of nondivergency of the rotational velocity field:

$$\frac{\partial \delta\rho}{\partial t} + \rho_0 \mathrm{div}\,\vec{v} = 0, \quad \mathrm{div}\,\vec{s} = 0. \qquad (2)$$

At the boundary of the rigid pore the translational velocity of the liquid vanishes and the rotational velocity is connected with the vortex through the linear expression [5]

$$\vec{v}(r,t) = 0, \quad \vec{s}(r,t) = \frac{\lambda}{2}\mathrm{rot}\,\vec{v}, \quad \lambda = \frac{\eta_r + \eta - \eta_b}{\eta_r}, \quad r = R, \qquad (3)$$

where $\eta_b$ is the viscosity at the interface between the two media (the liquid and the surface). The second boundary condition takes into account the difference between the bulk and surface viscosities [5].

The velocity and the random forces in Eqs. (1) can be divided into the longitudinal and transverse parts,

$$\vec{v} = \vec{v}_\parallel + \vec{v}_\perp, \quad \text{div}\,\vec{v}_\perp = 0, \quad \text{rot}\,\vec{v}_\parallel = 0. \tag{4}$$

As a result the equations for the transverse components of the translational velocity and the field of rotation separate from the rest of equations,

$$\rho_0 \frac{\partial \vec{v}_\perp}{\partial t} = -(\eta + \frac{\eta_r}{2})\text{rot}\,\text{rot}\,\vec{v}_\perp + \eta_r \text{rot}\,\vec{s}_\perp + \vec{f}_{v\perp}. \tag{5}$$

In spherical coordinates with the origin in the center of the pore the solutions of Eqs. (1) can be found using the expansion in the basis functions obeying the vector Helmholtz equations

$$\nabla \text{div}\,\vec{L} = -k^2 \vec{L}, \quad \text{rot}\,\text{rot}\,\vec{M} = k^2 \vec{M}, \quad \text{rot}\,\text{rot}\,\vec{N} = k^2 \vec{N}. \tag{6}$$

These systems of solutions should be taken to be finite at $r = 0$. They will be determined as the products of spherical harmonics and spherical Bessel functions [9,10],

$$L_{m,n}(\vec{r}) = \frac{1}{k} \nabla (Y_{mn}(\theta,\varphi) j_n(kr)), \tag{7}$$

$$M_{m,n}(\vec{r}) = \text{rot}(\vec{r} Y_{m,n}(\theta,\varphi) j_n(kr)), \quad N_{m,n}(\vec{r}) = \frac{1}{k}\text{rot}\,M_{m,n} = \frac{1}{k}\text{rot}\,\text{rot}(\vec{r} Y_{m,n}(\theta,\varphi) j_n(kr)).$$

The expansion for the translational and rotational velocity can be thus expressed as follows:

$$\vec{v}_\perp(\vec{r},t) = \sum_\lambda \left[ u_\lambda^M \widetilde{\vec{M}}_\lambda(\vec{r}) + u_\lambda^N \widetilde{\vec{N}}_\lambda(\vec{r}) \right], \quad \vec{v}_\parallel(\vec{r},t) = \sum_\lambda u_\lambda^L(t) \widetilde{\vec{L}}_\lambda(r), \tag{8}$$

$$\vec{s}_\perp(\vec{r},t) = \sum_\lambda \left[ s_\lambda^M \widetilde{\vec{M}}_\lambda(\vec{r}) + s_\lambda^N \widetilde{\vec{N}}_\lambda(\vec{r}) \right],$$

where we have introduced the normalized functions

$$\widetilde{\vec{a}}(\vec{r}) = \vec{a}(\vec{r}) / \|\vec{a}^2\|^{1/2}, \quad \|\vec{a}^2\| = \int_V d\vec{r}\, \vec{a}^2(\vec{r}).$$

From the conditions (3) it is easy to obtain transcendental equations that determine the eigenvalues $k_\lambda$ of the expansions of the fluctuating fields of translational and angular velocities,

$$j_n(\beta_{nl}) = 0, \quad k_\lambda = \beta_{nl}/R, \tag{9}$$

$$j_{n\pm 1}(\gamma_{nl}) = 0, \quad k_\lambda = \gamma_{nl}/R, \tag{10}$$

where Eq. (9) determines the eigenvalues for the vector function $\vec{M}$, and Eq. (10) the corresponding values for $\vec{L}$ и $\vec{N}$. The index $l$ is the root number and the collective index $\lambda$ stays for the set of three numbers $n$, $m$, and $l$. The normalizing factors are calculated taking into account the properties of the spherical harmonics and Bessel functions, and with allowance for Eqs. (7), (9), and (10) they take the form

$$\|\vec{a}_\lambda^2\| = q_\lambda \Lambda_\lambda^a, \quad q_\lambda = \frac{4\pi}{2-\delta_{m0}} \frac{1}{2n+1} \frac{(n-m)!}{(n+m)!}, \tag{11}$$

$$\Lambda_\lambda^M = \frac{R^3}{2} n(n+1) j_{n+1}^2(\beta_{nl}), \quad \Lambda_\lambda^L = R^3 \left[ \frac{1}{x} j_n(x) j_n'(x) + \frac{1}{2} j_n^2(x) \right],$$

$$\Lambda_\lambda^N = R^3 n(n+1) \left[ \frac{1}{x} j_n(x)(x j_n(x))' - \frac{1}{2} j_n^2(x) \right], \quad x = \gamma_{nl},$$

where $\delta_{ij}$ is the Kroneker symbol. We expand the longitudinal and transverse parts of the random forces $\vec{f}_v$ и $\vec{f}_s$ in the same way as in Eqs. (8), with the expansion coefficients $f_{i\lambda}^a$, where $a = L, M, N$ for $i = v$, and $a = M, N$ for $i = s$.

Rewriting Eqs. (1), (2), and (5) in the Fourier representation in time and taking into account the orthogonality of the eigenfunctions corresponding to different eigenvalues, we obtain algebraic equations for the Fourier components of the expansion coefficients,

$$u_\lambda^L = \frac{-i\omega f_{v\lambda}^L(\omega)/\rho_0}{-\omega^2 + c^2 k_\lambda^2 - i\omega k_\lambda^2 v_\parallel}, \quad v_\parallel = \frac{1}{\rho_0}\left(\frac{4}{3}\eta + \xi\right), \tag{12}$$

$$u_\lambda^{N,M} = \frac{1}{\rho_0} \frac{f_{v\lambda}^{N,M}(\omega)\left[-i\omega + \frac{1}{\tau} + D_s k_\lambda^2\right] + f_{s\lambda}^{M,N} \frac{1}{\tau} k}{-\omega^2 + k_\lambda^2 v_\perp \left[\frac{1}{\tau} + k_\lambda^2 D_s\left(1 + \frac{\eta_r}{2\eta}\right)\right] - i\omega\left[\frac{1}{\tau} + k_\lambda^2\left(D_s + v_\perp + \frac{\eta_r}{2\rho}\right)\right]}, \tag{13}$$

where $D_s = \mu/(\rho_0 I)$, $\tau = I\rho_0/\eta_r$, and $v_\perp = \eta/\rho_0$. Analogous expressions take place for the component of the rotational velocity.

The spectral densities of the expansion amplitudes of the fluctuating fields are found using the fluctuation-dissipation theorem [11]:

$$\langle \xi_j \xi_k \rangle_\omega = -\frac{ik_B T}{2\pi\omega}(\alpha_{jk} - \alpha_{jk}^*). \tag{14}$$

The spectral density $\langle \xi_j \xi_k^* \rangle_\omega$ is the Fourier transformation of the correlation function $\langle \xi_j(t) \xi_k^*(0) \rangle$ and $\alpha_{jk}$ is the generalized susceptibility matrix that relates the random fields $\xi_j$ to the corresponding generalized forces $f_k$:

$$\xi_j(\omega) = \sum_k \alpha_{jk} f_k(\omega). \tag{15}$$

The correspondence between the quantities $\xi_j$ and $f_k$ is found from the expression for the mean power dissipated in the system. The energy dissipated in the viscous compressible fluid with internal rotation can be written as follows [7,8]:

$$\frac{\partial E}{\partial t} = \int dV \left\{ \sigma_{\lambda\nu}^{os} (\nabla_\lambda v_\nu - s_\mu e_{\nu\mu\lambda}) + g_{\lambda\nu}^{os} \nabla_\lambda s_\nu \right\}. \tag{16}$$

Here $\sigma_{\lambda\nu}^{os}$ and $g_{\lambda\nu}^{os}$ are the external tensors of momentum and angular velocity flows. Equation (16) can be rewritten distinguishing the divergent contributions from the external sources. As a result the final expression for the energy dissipation is

$$\frac{\partial E}{\partial t} = -\int dV \left\{ \vec{v}\vec{f}_v^t - \left(\frac{1}{2}\text{rot}\vec{v} - \vec{s}\right)\vec{f}_v^a + \vec{s}\vec{f}_s^t - \frac{1}{2}\text{rot}\vec{s}\vec{f}_s^a \right\}, \tag{17}$$

where $f_{\nu\lambda}^t = \text{div}\,\sigma_{\lambda\nu}^{os,t}$, $f_{s\lambda}^t = \text{div}\,g_{\lambda\nu}^{os,t}$ are true vectors, $f_{\nu\lambda}^a = e_{\lambda\nu\mu}\sigma_{\lambda\nu}^{os,a}$, $f_{s\lambda}^a = e_{\lambda\nu\mu}\sigma_{\lambda\nu}^{os,a}$ are pseudovectors, and $e_{\lambda\nu\mu}$ is the Levi-Civitta tensor.

With the help of Eq. (17), the Langevin equations (13), and the fluctuation-dissipation theorem (14), the spectral densities of the thermal fluctuations of the expansion amplitudes of the hydrodynamic fields can be determined and then the spectral densities of the velocity correlation functions can be obtained. We shall give here the transverse component that is connected, as distinct from the longitudinal one, with the internal moment of the liquid inside the pore:

$$\langle \vec{v}_\perp(\vec{r},t)\vec{v}_\perp(\vec{r},t')\rangle_\omega = \frac{k_B T}{\pi\rho_0} \text{Re} \sum_{n>0,l} \frac{2n+1}{4\pi R^3} \tag{18}$$

$$\times \left\{ A(\beta_\lambda) \frac{2n(n+1)j_n^2(\beta_\lambda x)}{j_{n+1}^2(\beta_\lambda)} + 2A(\gamma_\lambda) \frac{n(n+1)[j_n(\gamma_\lambda x)/(\gamma_\lambda x)]^2 + \left[(\gamma_\lambda x j_n(\gamma_\lambda x))'/(\gamma_\lambda x)\right]^2}{j_n^2(\gamma_\lambda)\left[1 + \gamma_\lambda^{-2}(\pm(2n+1)+1)\right]} \right\},$$

$$A\left(\frac{k_\lambda}{R}\right) = \left[-i\omega + k_\lambda^2\left(\nu_\perp + \frac{\eta_r}{2\rho_0}\right) - \frac{1}{\tau}\frac{\eta_r k_\lambda^2}{2\rho_0}\left(-i\omega + \frac{1}{\tau} + D_s k_\lambda^2\right)^{-1}\right]^{-1}, \qquad x = r/R.$$

The signs "$\pm$" correspond to the signs in Eq. (10).

Using the result (18), the local coefficient of diffusion for the liquid inside the pore can be evaluated as follows:

$$D(r) = \frac{1}{3}\int_0^\infty \langle \vec{v}(\vec{r},t)\vec{v}(\vec{r},0)\rangle dt = \frac{1}{3}\langle \vec{v}(\vec{r},t)\vec{v}(\vec{r},0)\rangle_\omega. \tag{19}$$

The summation over the roots of Bessel functions is carried out up to the root not exceeding the value $a_{\max} = \pi R/a$, where $a$ is the interatomic distance, and the summation over the numbers of Bessel functions is from the first number to such a number $n_{\max}$, for which the first root of the Bessel function becomes larger than $a_{\max}$.

On Figure 1 some results of the calculations of the local diffusion coefficient are presented showing its dependence on the coordinate $r$ inside the pore of radius $R = 100$ Å. The rest of the parameters of the liquid are $D_s = 10^{-5}$ cm$^2$/s, $\tau = 10^{-10}$ s, $\rho_0 = 1$ g/cm$^3$, and $a = 5$ Å. Figure 2 demonstrates the frequency dependence of the autocorrelation function (normalized to its value at the zero frequency) of the transverse velocity component for various rotational viscosities $\eta_r$ at the distance $0.5\,R$ from the center of the pore. The parameters of the liquid are the same as above.

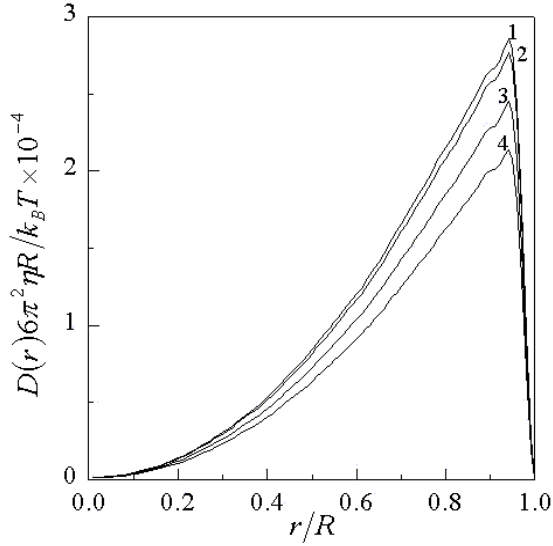 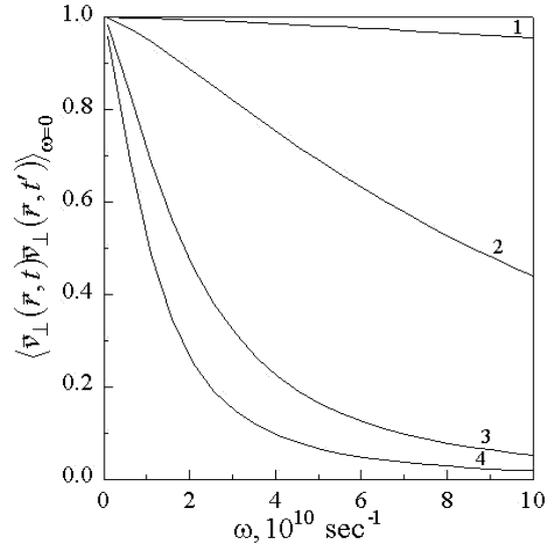

Figure 1. The local diffusion coefficient of a liquid inside the spherical pore: 1 – the liquid without an internal moment, 2 – the rotational viscosity is $\eta_r = 0.1$, 3 - $\eta_r = 0.5$, 4 - $\eta_r = 1$ poise.

Figure 2. The autocorrelation function of the transverse velocity component of a liquid inside the spherical pore at different rotational viscosities: 1- $\eta_r = 0.01$, 2 – $\eta_r = 0.1$, 3 - $\eta_r = 0.5$, 4 - $\eta_r = 1$ poise.

The presented theory generalizes the previously obtained results of the work [10] to the case of the fluctuations of liquids with internal moment in a pore. The confined space significantly influences the thermal hydrodynamic fluctuations of the compressible liquid. Besides this, the proper internal moment leads to the decrease of the local diffusion of the liquid.